\documentclass[color,superscriptaddress,floatfix,twocolumn]{revtex4}

\usepackage[dvips]{graphicx}
\usepackage{psfig}
\usepackage[latin1]{inputenc}
\usepackage{pstricks,pst-node,pst-text,pst-3d}
\usepackage{amsmath}
\usepackage{amsmath}
\usepackage{verbatim}
\usepackage{listings}
\usepackage{epic}
\usepackage{eepic}
\usepackage[dvips]{epsfig}
\usepackage[T1]{fontenc}

\begin{document}

\title{Numerical Methods as an Integrated Part of Physics Education}

\author{Arnt Inge Vistnes}

\affiliation{Department of Physics, 
University of Oslo, N-0316 Oslo, Norway}

\author{Morten Hjorth-Jensen}

\affiliation{Department of Physics and Center of Mathematics for Applications, 
University of Oslo, N-0316 Oslo, Norway}

\date{Final version \today}

\begin{abstract}

During the last decade we have witnessed an impressive development in so-called interpreted 
languages and computational environments such as Maple, Mathematica, IDL, Matlab etc. 
Problems which until recently were typically solved on mainframe machines and written in 
computing languages such as Fortran or C/C++, can now easily be solved on standard PCs with the 
bonus of immediate visualizations of the results.

In our undergraduate programs an often posed question is how to incorporate and exploit 
efficiently these advances in the standard physics and mathematics curriculum, without detracting 
the attention from the classical and basic theoretical and experimental topics to be covered. 
Furthermore, if students are trained to use such tools at early stages in their education, 
do such tools really enhance and improve the learning environment? And, perhaps even more important, 
does it lead to a better physics understanding? 

Here we present one possible approach, where computational topics are gradually baked into 
our undergraduate curriculum in Mathematics and Physics, Astronomy and Meteorology. 
We focus on training our students 
to use general programming tools in solving physics problems, in addition to the classical analytic problems. 
By this approach, the students gain an expertise that they can build upon in their 
future studies and careers. We use mainly Java, Matlab and Maple as computational environments. 
Our students are now capable of handling at an early stage in their education more realistic physics 
problems than before. We believe firmly that, in addition to educating modern scientists, 
this promotes a better physics understanding for a majority of the students.

\end{abstract}

\maketitle

\section{Introduction}

Computer simulations are nowadays an integral part of contemporary basic and applied research 
in the physical sciences.  Modern computational environments are widely used in industry as well. 
Computation is nowadays as important as theory and experiment. 
The ability "to compute" is now part of the essential repertoire of research scientists. 
Several new fields have emerged and strengthened their positions in the last years, 
such as computational materials science, bioinformatics, computational mathematics and mechanics, 
computational chemistry and physics, just to mention a few. 
Since computation enjoys such an important standing in the natural sciences and mathematics, 
we feel that our undergraduate curriculum should reflect this feature as well.

For this reason it seems obvious that university students in the physical sciences should get 
an education which reflects, in a coherent way (see the discussion below), 
how computers are used to solve problems in basic and applied research as well as in industry. 
Sadly, this is normally not the case. Such an education combines in principle knowledge 
from many different subjects, such as numerical analysis, computing languages and some 
knowledge of computers. These topics are, almost as a rule of thumb, taught in different, 
and we would like to add, disconnected courses. 
Only at the level of his/her thesis work is the student confronted with the synthesis 
of all these subjects. The usage of computers in solving problems in physics is 
often postponed to the master- or PhD-programs. In order to prepare better our students 
for their future careers and studies, we believe firmly that it is important to 
incorporate computational topics at an as early as possible stage in for example the 
undergraduate curriculum in physics. With the advances made in modern 
computational environments, which often include powerful tools to visualize immediately 
the results, we feel time is ripe for introducing such tools in the undergraduate curriculum. 

Furthermore, we believe that the introduction of numerical exercises can improve the 
learning environment and even add further physics insights to the more standard analytic exercises. 
By using interpreted languages and computational environments like Mathematica, Maple, Matlab etc, 
modeling often takes considerably less time than the more old-fashioned and traditional computing tools, 
which often are based on large Fortran and/or C/C++ programs and specialized visualization packages. 
Through our new educational university reform we have therefore started a project where the main aim 
is to gradually include computational exercises throughout the undergraduate studies in 
mathematics and physics. Programming languages such as Java and an interpreted language 
like Matlab are introduced as early as the  first and the second semester of the bachelor programs in 
Mathematics, Informatics and Technology (MIT) and Physics, Astronomy and Meteorology (FAM) at the 
University of Oslo. 
The usage and properties of these languages are added upon in later courses. 
At the end of the second semester the students are fully familiar with the syntax of for example Matlab and 
can use this environment professionally in more advanced undergraduate courses. 
This allows teachers in topics like electromagnetism or the introductory course in 
quantum physics to present more realistic problems, which hopefully convey further physics 
insights and make physics much more fun. At the end of their bachelor studies, 
our students are also exposed to other computational tools such as Maple or Mathematica 
and more traditional courses in computational sciences, courses which involve descriptions of 
algorithms and advanced problems solved with languages like C/C++.

The strategy for solving physics problems, which we have opted, is very close 
to the traditional one described by several textbook authors. 
Young and Freedman \cite{youngandfreedman} describe it as ISEE: Identify, Set up, Execute and Evaluate.
\begin{figure}
\includegraphics[width=0.95\columnwidth]{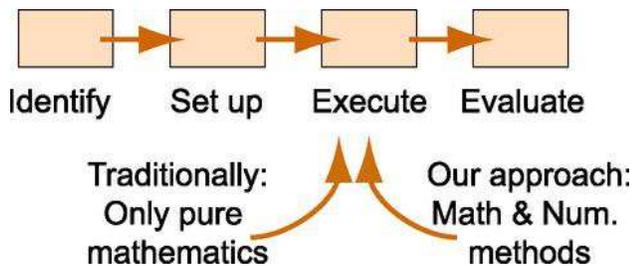}
\caption{Strategy for solving physics problems, and pointing out where numerical solutions come into play.}
\end{figure}
The numerical methods only replace some of the pure analytical tasks that make up the 
Execute part of this diagram. Our approach is an algorithmic one, which stresses the 
understanding of the underlying physics. This is also the reason we prefer tools 
like Matlab or Maple, where the students, under guidance, have to model the actual physical systems. 
This approach should be contrasted to the widespread use of applets to demonstrate physics problems.

Many people think computers already are included in the physics education. 
In one way this is correct. Thousands of applets have been written for physics alone, 
and by these applets students may for example visualize the path of a ball through air. These applets 
may be useful as a pedagogical tool, hopefully giving the students a better understanding of various phenomena. 
However, the usage of  applets made by others do not necessarily increase 
the students ability to solve physics problems in general. An applet made by others is 
very seldom to much use when the student is to solve physics problems in a later job. 
Furthermore, it is often difficult to get a close relationship to an applet made by others, 
since one often feels insecure about what is actually going on below the surface of the applet, 
that is, how the modeling in fact is done. In our experience, use of applets, for these reasons, 
has a more limited importance than one would expect.

In the next section we present the basic structure of our program on 
'Computers in Science Education' applied to the bachelor programs FAM and MIT, with an emphasis
on the Physics program. Sec.~\ref{sec:sec3} presents examples 
of numerical exercises the students are exposed to during their two first years of physics studies.
Finally, we present our summaries and perspectives in Sec.~\ref{sec:sec4}.

\section{Computers in Science Education}\label{sec:sec2}

We started to implement numerical exercises in our undergraduate courses back 
in 1999. The first few years we implemented the new tools in two or three mathematics and physics courses only, 
with mixed experiences. At that time many undergraduate students were not familiar with 
programming concepts in the natural sciences, except for an introductory (not mandatory) programming course, 
where typically some high-level language like Java was taught.  The students 
which attended this course would perhaps not meet computational problems before they embarked on 
their master or PhD thesis, with typical time spans from when they learned computing till 
they started on a thesis project of three to four years. The emphasis of this programming course 
was and still is on general concepts, with few, if any, applications to problems from 
the physical sciences. Another important problem was and still is that of the readiness of our teachers. 
Very few were and still are familiar with the new programming tools and could therefore not aid in an efficient 
way our students in solving problems. Finally, the infrastructure with computer labs, software, 
operating systems etc was not optimal, or it was not clear which solution to choose, 
for large classes of students. 

Since then we have benefited from several advances. We have invested in instruction labs with computers,  
in addition we 
have general computer rooms that are open most of the day. We also have wireless networks in all common 
rooms so that students can link their laptops to the local network and get access to different 
software packages, course pages with exercises, lecture notes etc. Both 
Linux and Windows environments are available, and there is some support for Mac OS. 
Student versions at low prices exist for many of the interpreted languages.

However, in spite of the above improvements we feel that three 
basic problems still persist. These are
\begin{itemize}
\item How to gradually bring in computational aspects in most of our undergraduate courses 
without taking away the attention from the basic topics which have to be covered.
The standard complaint from many university teachers is that computational exercises 
bring in an additional complication to an already difficult topic and they do not 
wish to spend time on it during the regular classes.
\item Most teachers are not familiar with many of the new programming tools.
A dedicated program to increase the level of knowledge of our teachers is 
needed for this program to be successful.
\item Finally, to make meaningful exercises which go beyond the available body of 
very good analytic exercises and can trigger further insights represents perhaps the greatest challenge
we face.
\end{itemize}

Through our recent university reform dating from the fall 2003 and an extensive collaboration 
with the bachelor program in Mathematics (MIT), we have now established a project called 
'Computers in Science Education', where the aim is to address the above three topics.  
The reform of 2003 introduced a three-year bachelor program, a two-year master program and 
finally a three-year PhD program. The introduction of the bachelor program allowed us to re-model the 
undergraduate curriculum, implying a large degree of common and compulsory courses in physics, 
informatics and mathematics. Our two bachelor programs have typically 200-300 new students 
every year, with roughly 100 starting with Physics, Astronomy and Meteorology.

Our physics and mathematics students follow the same courses the first semester. 
The first semester consists of a standard mathematics course on analysis and calculus, 
one on programming (Java) and finally a course on modeling and mathematics applications. 
The latter two courses introduce programming concepts using Java, operating systems such as Linux and Windows, 
use of editors, numerical algorithms for integration and solution of differential equations, 
representations of numbers, roots of transcendental equations and so forth. 
This gives our students a common and uniform background in various tools and programming environments. 
Our students meet Matlab applications the second semester through the mechanics course 
(see section \ref{sec:sec3}), a mathematical course on field theory and vector analysis 
and a course on linear algebra. The latter two courses are also part of the bachelor program 
in mathematics whereas the mechanics course is optional for mathematicians\footnote{Our fall semester 
lasts 19 weeks,
while the spring semester lasts 21 weeks. The last two to three weeks of every semester are reserved
for the final exams. On average 15-16 weeks of each semester are dedicated to regular teaching.
Every course gives ten credits in the new ECTS system. }

Our students are thus exposed to programming topics and numerical exercises in five out of six 
undergraduate courses the first year of study. This allows us to portion out the 
computational learning threshold over time and in different courses, 
avoiding thereby much of the well-known criticisms discussed above.

With this caveat we are able to give our students a uniform background in computational skills, 
skills which can then be used in solving slightly more advanced problems, without taking away 
the attention from the interesting physics. The second year of study begins then 
with a course on electromagnetism, an optional course on astrophysics/meteorology 
and an advanced linear algebra and calculus course. The electromagnetism course contains several 
numerical exercises, including even solutions of partial differential equations such as Laplace's equation. 
Matlab is used in solving such problems. The second year ends with the spring semester and 
with an introductory course on quantum physics, a course on wave equations and oscillations 
and an experimental physics course. All these courses offer numerical exercises 
and use computers for project writing. At this stage most students are fairly familiar 
with computational topics. The last year of the bachelor degree contains more advanced topics, 
including topics such as statistical physics, mathematical physics, computational physics, solid state physics, 
subatomic physics, advanced quantum mechanics and so forth. Here it is up to the various teachers whether 
they wish to include numerical exercises or not. However, the background of the students for doing this 
is now much better 
than previously. 

To meet the needs of our university teachers we have developed and arrange 
on a regular basis intensive one week courses on various computer tools. We have  developed courses on 
Matlab, Maple, visualization tools, Fortran95, C/C++, Scripting languages and high-performance computing. 
All course material is available on the net for self-study, with a large body of exercises. For links to these
courses see Ref.~\cite{courselinks}. Most of the material is in english.

Below we discuss two examples of physics exercises which can be solved using Matlab.

\section{Examples from Physics Courses}\label{sec:sec3}

We had several aims in mind when we embarked on computational exercises in our undergraduate physics courses. 
\begin{itemize}
\item We wanted to give the students an opportunity to gain a deeper understanding of the physics. 
In most courses one is normally confronted with simple systems which provide exact solutions and mimic 
to a certain extent the realistic cases. Many are however the comments like 'why can't we do something 
else than the classical box potential in quantum mechanics?' In several of the numerical exercises 
we propose more 'realistic' cases to solve by various numerical methods. This also means that 
we wish to give examples of how physics can be applied in a much broader context than it is discussed 
in the traditional physics undergraduate curriculum. 
\item To encourage the students to "discover" physics in a way similar to how researchers learn in the 
context of research.
\end{itemize}
Our overall goal is to encourage the students to learn science through experience and by asking questions. 
Our objective is always a deeper understanding.
The purpose of computing is further insight, not mere numbers! 
Moreover, and this is our personal bias, to device an algorithm and thereafter write a code 
for solving physics problems is a marvelous way of gaining insight into complicated physical systems. 
The algorithm one ends up writing reflects in essentially all cases the understanding 
of the physics of the problem. 

Below we discuss two simple applications which illustrate the above aims, one from our mechanics 
course (second semester of the first year of study) and the other 
from our introductory course in quantum physics (second semester of the second year of study).

\subsection{Mechanics Problems}

\begin{figure}
\includegraphics[width=0.95\columnwidth]{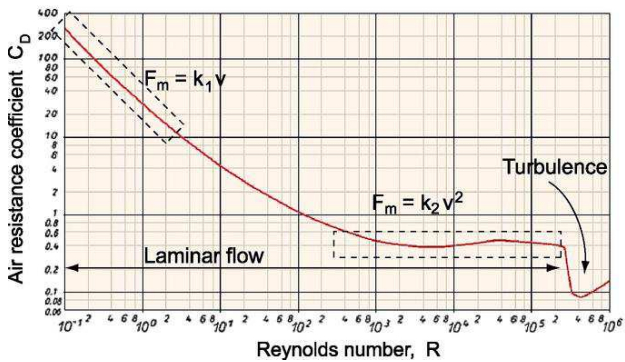}
\caption{Air resistance vs. Reynold's number for a sphere moving through air. 
See Physics World for similar details (http://physicsweb.org/articles/world/11/6/8/1/world-11-6-8-3).}
\label{fig:fig2}
\end{figure}

It would have been practically impossible to treat a realistic projectile motion 
for a football by using the traditional analytical tools alone. The reason is that 
the air resistance varies in a rather complicated way for the football problem.  
Figure \ref{fig:fig2} displays the air resistance coefficient for a spherical 
ball moving through air (for corresponding equations, see below). The initial slope 
for low speeds makes the air resistance approximately proportional with the speed. 
In the plateau region of the curve, the air resistance is roughly proportional with the square 
of the speed. However, for really high ball speeds, which professional football players actually 
obtain, the air resistance drops drastically. 

Since there are no single purely analytical expression with constant coefficients that can be 
used to describe the air-resistance throughout the projectile motion for a football, 
it is impossible to work out a solution in closed form. However, when we solve the projectile 
motion by numerical methods, only approximately five to ten extra lines of code are needed in 
order to include all the details of Figure 2 in the calculations. The full Matlab code 
for this problem, implementing Euler's method for solving a second order differential equation, 
and plotting of the results, can easily be written with 50 lines of code. 
This represents a manageable problem for our students. Once such a program is written, 
it can easily be modified to quite different problems. 
With a few changes it can for example handle a realistic rocket launch 
(for scientific rockets that our department sends up in the atmosphere in order to study auroras). 
In the rocket problem almost every thinkable parameter changes during 
the flight (for example, the rocket engine thrust, the air resistance and even the gravitation force).

The expression for the air resistance we use is given by the following:
\begin{equation}
         {\bf F} = - \frac{1}{2}\rho S C_d v{\bf v},
\end{equation}
where $\rho$ is the density of the air, $S$ is the cross section 
of the ball, $C_d$ is the air resistance coefficient, $v$ is the speed, and ${\bf
v}$ is the velocity of the ball (a vector). It is the variation of the 
coefficient $C_d$ that is displayed in Figure 2. The Reynolds number is proportional with the velocity, as usual.

When using numerical methods, focus tends to be on Newton's second law, not on for example the 
final integrated solutions for the simplified system without air resistance. 
Furthermore, the students realize that the Execute part of Figure 1, is merely a 
tool part that is rather separate from the physics involved. In the traditional approach 
the mathematics part (execute part) is often focused on to such a degree that it is common 
to think that that part is the important one in the problem solving, whereas the opposite often is the case.
Thus, by introducing numerical methods, the students can more easily grasp what is the important physics 
in the problem, and what is not. And again, 
the realism of the problem helps to inspire the students in their work. 

On the other hand, our approach is more demanding than the traditional one. 
Of course it is easier to use blindly the final equations for a projectile motion 
without air resistance, than to solve a more realistic problem, since in the latter case 
the students must go somewhat deeper into the problem. Our approach therefore tends to some degree 
to exclude our weakest students.  This represents a challenge to your physics education.

The above is just one of many examples of physics exercises included in our undergraduate courses. 
Other numerical exercises included in for example our mechanics (second semester) and 
electromagnetism (third semester) courses are
\begin{enumerate}
\item Hunting gun bullet (ballistic) movement with complex description of air
resistance.
\item Rocket launching with almost all realistic parameters.
\item Mathematical and physical pendulum with large amplitudes.
\item Chaotic motion.
\item Planetary movement, position of planets compared to the background star sky.
\item Calculate magnetic field from arbitrary current geometries, based on Biot-Savart's law.
\item Transients when a forced harmonic oscillation starts, for different Q-factors.
\item Different forms for electromagnetic waves, animations.
\item Combined effects of various potentials (e.g. electrostatic and gravitational).
\end{enumerate}

\subsection{Simple Examples from Quantum Physics}
In the traditional first course in quantum physics (which at our university is taught towards the end of 
the second year), students are exposed to topics like one-dimensional potential problems represented by the
harmonic oscillator, the standard square well potential and the hydrogen atom. In addition, the students get  
a basic understanding of the 
periodic system, molecular physics, some nuclear, particle and solid state physics. The more formal mathematical
framework of quantum mechanics is presented in a  senior undergraduate course in the fifth
semester of study (beginning of the last year of the bachelor degree). 

\begin{figure}[h]
\begin{center}
\setlength{\unitlength}{0.7cm}
\begin{picture}(13,6)
\thicklines
   \put(0,0.5){\makebox(0,0)[bl]{
              \put(8,1){\vector(1,0){4}}
              \put(12.3,1){\makebox(0,0){x}}
              \put(5.1,1.5){\makebox(0,0){$-a$}}
              \put(8.1,1.5){\makebox(0,0){$a$}}
              \put(4,0){\makebox(0,0){$I$}}
              \put(6,0){\makebox(0,0){$II$}}
              \put(10,0){\makebox(0,0){$III$}}
              \put(8.5,-3){\makebox(0,0){$-V_0$}}
              \put(5,1){\line(0,-1){4}}
              \put(5,1){\line(-1,0){3}}
              \put(5,-3){\line(1,0){3}}
              \put(8,1){\line(0,-1){4}}
         }}
\end{picture}
\end{center}
\caption{Simple quantum mechanical problem for a particle confined in a square well potential.}
\label{fig:fig3}
\end{figure}
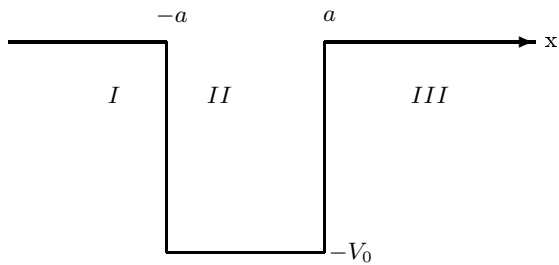

A standard textbook exercise 
the students could be exposed to is to solve Schr\"odinger's equation for a particle confined
in a one-dimensional square well potential, given by for example Figure \ref{fig:fig3}.
This exercise cannot be solved analytically.
The one-dimensional  Schr\"odinger  equation takes the following form 
\begin{equation}
-\frac{\hbar^2}{2m}\frac{\partial^2}{\partial x^2}\psi(x) +
V(x)\psi(x) = -|E|\psi(x),
\end{equation}
and can be rewritten
\begin{equation}
\frac{\partial^2}{\partial x^2}\psi(x) - \frac{2m}{\hbar^2}\left( V(x)
+ |E| \right) \psi(x) = 0.
\end{equation}
Note that $V(x) + |E| \leq 0$. For regions $I$ and $III$ the potential is zero and we have
\begin{equation}
\frac{\partial^2}{\partial x^2}\psi(x)-\beta^2\psi(x)=0,
\end{equation}
with $\beta^2=2m|E|/\hbar^2$. In region $II$  $V(x)=-V_0$, resulting in
\begin{equation}
\frac{\partial^2}{\partial x^2}\psi(x)+k^2\psi(x)=0,
\end{equation}
where $k^2=2m(V_0-|E|)/\hbar^2$ is real and positive.
In all three regions we have simple second-order differential equations
with their pertinent boundary equations.

The standard exercise the students are exposed to (and we claim this is done in an almost uniform
way all over the world) is to set up Schr\"odinger's equation for these three regions and show that
one can obtain a set of transcendental equations for the energy and thereby the wave functions.
They would typically proceed by setting up the general solutions for 
regions $I$ and $III$ as follows 
\[ \psi_I(x) = Ae^{\beta x} + Be^{-\beta x} \]
and
\[ \psi_{III}(x) = Fe^{\beta x} + Ge^{-\beta x}. \]
Similarly, they would show that for region $II$ one arrives at
\[ \psi_{II}(x) = C\cos(kx) + D\sin(kx). \]
The coefficients 
$A$, $B$, $C$, $D$, $F$ and $G$ are all unknown. 

The students have to distinguish between symmetric and antisymmetric solutions,
meaning that the wave functions obey either $\psi(x)=\psi(-x)$ or
$\psi(x)=-\psi(-x)$. 
This gives 
$D=0$ for the symmetric case and 
$C=0$ for the antisymmetric case. In addition, the wave functions have to be bounded.

The final step is to use the requirement that the wave function is continuous at $x=a$ and
$x=-a$ and after some minor algebra the students arrive at 
\begin{eqnarray}
&\psi_{III}(a)=\psi_{II}(a) \quad = \quad Ge^{-\beta a} = C\cos(ka)&\\
&\dot{\psi_{III}}(a) = \dot{\psi_{II}}(a) \quad = \quad -\beta Ge^{-\beta a} =
-kC\sin(ka),
\end{eqnarray}
 resulting in the the following transcendatal equations
\begin{equation}
 \beta = k \tan(ka),
\end{equation}
for the symmetric solution and
\begin{equation}
 -\beta = k \cot(ka), 
\end{equation}
for the antisymmetric case.

In the traditional approach, this is normally where one would stop. The emphasis is then obviously only on the
mathematical manipulations in order to arrive at the two final equations. One could always plug in 
values for the potential and its range resulting in analytical functions, however this limits very much
the physics investigations which can be made with modern tools.

As an example, Matlab, Maple and Mathematica all have functions for solving numerically the above 
non-linear equations and the students can easily investigate the number of solutions as function
of the depth of the potential and its range. They can also write their own algorithms for solving
such equations.
The next step is to use the eigenvalues to obtain the wave functions and plot these as function of the 
potential depth and the range. A useful exercise is to study the behavior of the wave 
function 
as the range of the potential increases. The
oscillating wave function one then sees from the plot  corresponds to the case of a free particle.  

These problems can easily be solved with most of the above tools, providing the students with an extremely
useful tool for asking further physics questions.
The simplest way to implement the above is then as follows
\begin{enumerate}
\item Use the functions of Matlab, Maple or Mathematica for solving non-linear equations. Obtain the eigenvalues
as function of the depth and the range of the potential.
\item Write a small code which includes these solutions and obtains the eigenfunctions for
the symmetric and the antisymmetric case. Plot these functions and compare to the free particle case.
\end{enumerate}  
Most students are capable of implementing the above two steps. There is no need to consider 
the solution of differential equations with boundary problems and our experience is that the students
find the immediate visualization of the results as very rewarding and interesting. 
Such a code is however not extendable to more interesting cases such as the hydrogen atom or the harmonic
oscillator. 
One can increase the degree of numerical complexity by letting the students write more generic codes for
single-particle problems. Matlab offers a set of functions for solving differential equations with boundary 
conditions. This is exposed in the following Matlab code for the particle in a square well potential.
\begin{lstlisting}[title={Matlab code for the particle in a square well  potential}]
function BoxPotential
clear
 
% Global constants (units: nm and eV) 
global m hbarc V0 a factor; 
m=511000.;    	% electron mass [eV/c2] 
hbarc=197.327; 	% hbar*c [eV nm]
V0=5.;       	% Potensial depth [eV] 
a=0.1;       	% Extension of V [nm]
factor = 2*m/hbarc/hbarc; 	% factor in front of  (V0-|E|)
 
%   Initial guess for the energy
lambda = 2.;

solinit = bvpinit(linspace(0,0.8),@matBOXinit,lambda)
 
% BVP4C returns the  structure 'sol'. The eigenvalue is
% in sol.parameters.  The mesh points used by Matlab
% are in sol.x, while y(x) is in sol.y 
% sol.yp contains the derivative of sol.y using sol.x
 
sol = bvp4c(@matBOXode,@matBOXbc,solinit);
fprintf('Eigenvalue is %7.3f.\n',sol.parameters)
 
% Here we define an array of x-values for 
% where to plot y
xint = linspace(0,0.8);
% Thereafter we interpolate to get  
% the solution y(xint(i))

Sxint = deval(sol,xint);
% prepare norm of wf
% squared wf
wf2 = Sxint.*Sxint;
% Norm by trapezoidal rule
hint = trapz(xint,wf2(1,:));
% Finale wf
Sxint = Sxint/sqrt(hint);
% plot the wave function
figure;
plot(xint,Sxint(1,:));
axis([0 0.8 -5 5]);
title('Eigenfunction');
xlabel('x');
ylabel('Solution y');
 
% The potential
 
function v = Boxpot(x)
global V0 a
if ( x < a)
    v = V0;
elseif ( x >= a)
    v = 0;
end
 
% The ODE, two columns: y(1) and  y(2) and dydx(1) and dydx(2)
function dydx = matBOXode(x,y,lambda)
global factor
dydx = [              y(2)
         -factor*(Boxpot(x)-lambda)*y(1) ];
 
% Boundary conditions
function res = matBOXbc(ya,yb,lambda)
res = [  ya(2)
         yb(1)
         ya(1)-1];
 
% trial for the solution and its derivative
function yinit = matBOXinit(x)
yinit = [   cos(x)
    -sin(x)];
\end{lstlisting}
The above is for a particle moving in a square well potential, however it is easy to extend this
to other types of single-particle potentials. It allows, by simply changing the differential equations,
the boundary conditions and the potential, to attack many other single-particle problems and provides
thereby a generic background which is independent of the specific physics problem. This is 
obviously in line with modern computing concepts which our students are exposed to in courses
on programming concepts. 
As such it allows one to focus on the basic physics, represented by the boundary conditions and the potential.
This abstraction is however difficult for most students, and a balance has to been sought. 
It is also a much more difficult procedure to grasp 
as it hides the solution of a differential equation
with boundary conditions. 
On average, the students find the usage of the above approach rather difficult since the way it is 
coded is fairly compact. By presenting this algorithm here we also expose a difficulty inherent to most
modern languages like Maple or Matlab. The coding of the physics represented by the boundary conditions
and the potential is highly non-trivial and non-transparent. 
It would be much better if the students could code directly the boundary conditions, the potential and the 
differential equation to be solved in a way which is close to the mathematics.

In the fourth semester most students are not familiar with such problems and we do no
recommend to use such an approach.
The students need to provide the boundary conditions on the wave function and its derivative based on the symmetry 
of the solution, they need
to code the potential itself and the form of the second-order differential equation. 
Our experience when we used the above more generic approach, is that it is actually 
better to let the students code explicitely the shooting method for finding the eigenvalues.
This is however more time-demanding than the steps outlined prior to the code above.

\section{Summary}\label{sec:sec4}
The new approach in physics education at our university has been rather successful as far as we can judge. 
We can summarize the benefts and problems as follows:

Benefits:
\begin{itemize}
\item The students meet more interesting and realistic problems than before
\item We believe this helps in motivating and inspiring our students to pursue a physics career. 
\item The choice between traditional analytical mathematical tools alone and the same 
tools combined with numerical tools, clarifies in a better way what is the important physics and which are
the useful tools to be used in  problem solving.
\item We can devote more time on the core physics equations.
\item The students get an expertise 
that can make them more productive in their future jobs and careers.
\item It gives the university teachers a greater advantage since it brings the students closer to the 
way we do research. It is more fun to teach. Our experience is that we gain much more insights ourselves
than in the more traditional approaches since we can address much more interesting questions.
\end{itemize}
Problems:
\begin{itemize}
\item Many university teachers do not know how to use the new tools. We offer therefore several
intensive courses in order to upgrade the knowledge of our staff. 
\item However, there will always be a certain number of teachers  who 
do not like to spend time on the new tools since that leaves less time 
for traditional teaching and /or student activities. Parts of this reluctance is due to pure inertia, 
but most of the criticism  is very actual, and care has to be exercised so that 
the time spent on numerical methods does  not become too large 
compared with the total activity.
\item We have to give more individual instructions and  feedback to the students than before. 
We end up using more teaching resources.
\item Very few textbooks offer problems where numerical methods are required for obtaining a solution, thus: 
        - We have to develop new problems ourselves. It takes time but it is definitely worth it.
\item Some of the weaker students have problems in getting through the new courses, 
whereas the clever students enjoy the flexibility that the new tools give them. This represents a clear 
pedagogical 
challenge to our physics education.
\end{itemize}
Fortunately, many of the problems will be  reduced as time passes since students get more used to
computational exercises and we make (hopefully) better physics problems.

\end{document}